\documentclass[floats,floatfix,showpacs,amssymb,prd,twocolumn,superscriptaddress,nofootinbib,nolongbibliography,reprint]{revtex4-1}

\usepackage{amssymb,amsmath,verbatim,mathtools,needspace,enumitem,etoolbox,graphicx,physics,microtype,afterpage,xspace,tabularx,lmodern,multirow}
\usepackage{gensymb}

\usepackage[dvipsnames, usenames]{xcolor}

\definecolor{linkcolor}{rgb}{0.0,0.3,0.5}
\usepackage[unicode, colorlinks=true, linkcolor=linkcolor, citecolor=linkcolor, filecolor=linkcolor,urlcolor=linkcolor, pdfusetitle]{hyperref}
\usepackage[all]{hypcap}
\usepackage[T1]{fontenc}
\usepackage[utf8]{inputenc}
\usepackage[usenames,dvipsnames]{xcolor}
\newcommand{\ssim}{\mathchar"5218\relax\,}

\newcommand{\jhu}{\affiliation{Department of Physics and Astronomy, Johns Hopkins University, 3400 N. Charles St, Baltimore, MD 21218, USA}}
\newcommand{\bham}{\affiliation{School of Physics and Astronomy \& Institute for Gravitational Wave Astronomy, \\ University of Birmingham, Birmingham, B15 2TT, UK}}
\newcommand{\ligo}{\affiliation{LIGO Laboratory, Massachusetts Institute of Technology, 185 Albany St, Cambridge, MA 02139, USA}}
\newcommand{\mitt}{\affiliation{Department of Physics \& Kavli Institute for Astrophysics and Space Research, Massachusetts Institute of Technology, 77 Massachusetts Ave, Cambridge, MA 02139, USA}}

\def\araa{\rm{ARA\&A}}
\def\apj{\rm{ApJ}}
\def\apjl{\rm{ApJ}}
\def\apjs{\rm{ApJS}}
\def\aap{\rm{A\&A}}

\def\mnras{\rm{MNRAS}}
\def\prd{\rm{PRD}}
\def\prl{\rm{PRL}}
\def\prx{\rm{PRX}}

\def\cqg{\rm{CQG}}
\def\lrr{\rm{LRR}}

\def\chieff{\ensuremath{\chi_{\rm eff}}\xspace}
\def\chip{\ensuremath{\chi_{\rm p}}\xspace}

\newcommand{\event}{GW190412\xspace}

\newcommand\prlsec[1]{\vspace{2mm}\noindent {\bf \emph{#1}}}

\newcommand\orcid[1]{\href{https://orcid.org/#1}{$\!\!$\includegraphics[scale=0.006]{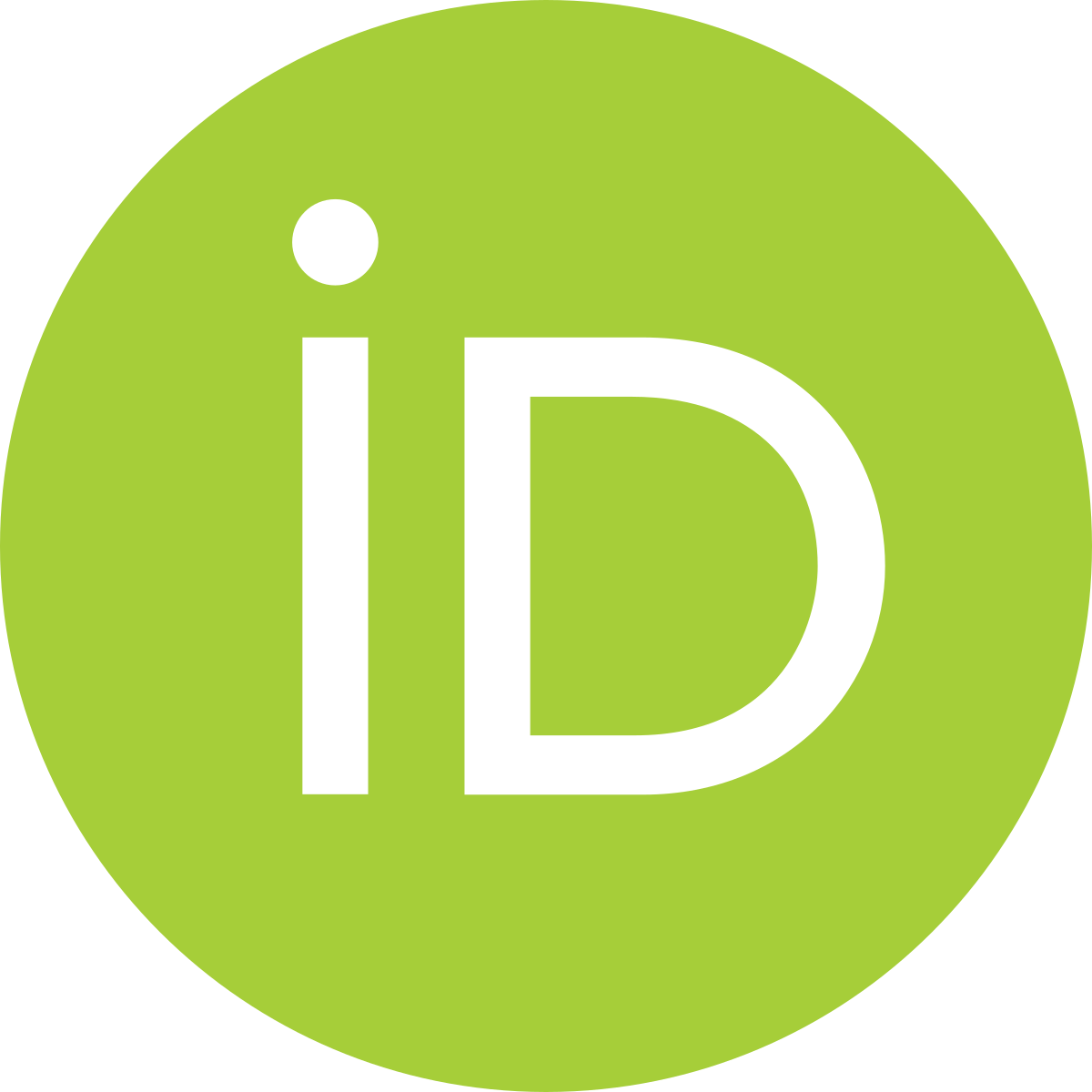} $\!\!$}}

\begin{document}

\title{Astrophysical implications of \event as a remnant of a previous black-hole merger}

\author{Davide Gerosa \orcid{0000-0002-0933-3579}}
\email{d.gerosa@bham.ac.uk}
\bham
\author{Salvatore Vitale \orcid{0000-0003-2700-0767}}
\email{salvatore.vitale@ligo.mit.edu}
\ligo
\mitt
\author{Emanuele Berti \orcid{0000-0003-0751-5130}}
\email{berti@jhu.edu}
\jhu
\pacs{}
\date{\today}

\begin{abstract}
Two of the dominant channels to produce merging stellar-mass black-hole binaries are believed to be the isolated evolution of binary stars in the field and dynamical formation in star clusters. The first reported black-hole binary event from the third LIGO/Virgo observing run (\event) is unusual in that it has unequal masses, nonzero effective spin, and nonzero primary spin at 90\% confidence interval. We show that this event should be exceedingly rare in the context of both the field and cluster formation scenarios. Interpreting \event as a remnant of a previous black-hole merger provides a promising route to explain its features. If \event indeed formed hierarchically, we show that the region of the parameter space that is best motivated from an astrophysical standpoint (low natal spins and light clusters) cannot accommodate the observation.  We analyze public \event LIGO/Virgo data with a Bayesian prior where the more massive black hole resulted from a previous merger, and 
find that this interpretation is equally supported by the data.
If the heavier component of \event is indeed a merger remnant, then its spin magnitude is $\chi_1=0.56_{-0.21}^{+0.19}$, which is higher than the value previously reported by the LIGO/Virgo collaboration.
\end{abstract}

\maketitle

\prlsec{Introduction.}
The first observation of a merging black-hole (BH) binary reported from LIGO/Virgo's \cite{2015CQGra..32g4001L,2015CQGra..32b4001A} third observing run (O3), \event,  is %
unusual in many ways~\cite{2020arXiv200408342T}.

A BH binary is characterized by component masses $m_1$ and $m_2$, and spins with dimensionless magnitudes $\chi_1$ and $\chi_2$. The masses are conveniently combined into
chirp mass $M_c = (m_1 m_2)^{3/5} / (m_1+m_2)^{1/5}$ and mass ratio $q=m_2/m_1\leq 1$. The LIGO/Virgo collaboration reports spin constraints in terms of the effective spin %
$\chi_{\rm eff} = (\chi_1 \cos\theta_1 + q \chi_2 \cos\theta_2)/(1+q)$~\cite{2001PhRvD..64l4013D}
and of the precession parameter
$\chi_{\rm p} =  \max [ \chi_1\sin \theta_1,\, \chi_2 \sin\theta_2  q(4q+3)/(4+3q) ]$~\cite{2015PhRvD..91b4043S},
where $\theta_{1,2}$ are the angles between the individual BH spins and the orbital angular momentum of the binary.

The events detected in the previous LIGO/Virgo observing runs (O1 and O2) had nearly equal masses ($q\simeq 1$), aligned spins components resulting in $\chieff\simeq 0$, and did not allow for meaningful measurements of $\chip$~\cite{2019PhRvX...9c1040A,2019ApJ...882L..24A} (but see~\cite{2019arXiv191009528Z,2020arXiv200304513H}).
The event \event~\cite{2020arXiv200408342T}
is unusual because it has unequal masses ($q=0.28_{-0.07}^{+0.13}$), the
effective spin $\chi_{\rm eff}= 0.25_{-0.11}^{+0.09}$ is nonzero, and there is marginal evidence for spin precession ($\chi_{\rm p}= 0.30_{-0.15}^{+0.19}$).  The source-frame chirp mass is $M_{c}= 13.27_{-0.32}^{+0.40} M_\odot$. %
 Here we quote medians and 90\% symmetric credible intervals obtained by combining samples from different waveform families, although there are some systematic differences between the models~\cite{2020arXiv200408342T}.

Based on the population of BHs detected during O1 and O2, Ref.~\cite{2020ApJ...891L..31F} predicted that 99\% of the events should have $q\gtrsim 0.5$, and \event is well outside that region. The unusual character of this event is also reflected in the population fit reported in Ref.~\cite{2020arXiv200408342T}, where the inferred slope of the mass-ratio spectral index is found to change dramatically (from $\beta\ssim 7$ to $\beta\ssim 0$) when \event is included in the population. However, this conclusion is questionable. \event was chosen from $\ssim 50$ O3 triggers~\cite{GraceDB-O3-Public} and analyzed with priority precisely because of its unusual properties. Therefore, combining it with the previous limited sample of only ten events from O1 and O2 can produce statistical biases. %

Astrophysical models of core-envelope interactions in massive stars predict that most BHs are born very slowly rotating ($\chi\ssim 0.01$)~\cite{2019ApJ...881L...1F}.  Ref.~\cite{2020ApJ...895..128M} found that BH effective spins from O1 and O2 are indeed distributed around zero with a dispersion $\lesssim 0.1$, and this can have important implications in terms of population inference~\cite{2020PhRvD.102d3002B,2020arXiv200500023K}.
With a measured primary spin $\chi_1\ssim 0.43$ \cite{2020arXiv200408342T}, \event challenges previous predictions. Indeed, its unusual properties have already sparked numerous interpretations in the astrophysics community, ranging from isolated binaries with tidally spun-up secondaries \cite{2020ApJ...895L..28M,2020arXiv200411866O,2020ApJ...897L...7S} to dynamical assembly in young star clusters \cite{2020arXiv200409525D}, gas-assisted migration in the disks of active galactic nuclei (AGN)~\cite{2020ApJ...899...26T}, quadruple stars~\cite{2020ApJ...898...99H}, and super star clusters~\cite{2020ApJ...896L..10R}.

Our main goal is to investigate whether one of the components of \event can be interpreted as the remnant of a previous BH merger. Remnant BHs left behind following mergers present a characteristic spin distribution peaked at $\chi\ssim 0.7$~\cite{2008ApJ...684..822B,2017PhRvD..95l4046G,2017ApJ...840L..24F}, thus providing a natural way to reconcile the measured value of $\chi_{\rm eff}$ with small natal spins. Because  remnants are, on average, more massive than BHs originating from the collapse of stars, this might also explain the low value of $q$.
However, linear momentum dissipation during the late inspiral and merger imparts a recoil to the BH remnant. %
A hierarchical merger interpretation is viable only in an environment that (i) allows for frequent dynamical interactions such that BH remnants can merge again but, at the same time, (ii) prevents the ejection of merger remnants due to gravitational recoils~\cite{2019PhRvD.100d1301G}.  Point (i) excludes isolated binary formation, while point (ii) excludes light dynamical environments such as globular clusters and young star clusters. These have escape speeds $\lesssim 50$\,km/s~\cite{2004ApJ...607L...9M,2010ARA&A..48..431P}, while typical gravitational recoils are $\mathcal{O}(100)$\,km/s~\cite{2010ApJ...715.1006K,2012PhRvD..85l4049B,2012PhRvD..85h4015L,2018PhRvD..97j4049G}. Environments with larger escape speeds, like AGN disks~\cite{2017MNRAS.464..946S,2017ApJ...835..165B,2018MNRAS.474.5672L} and/or nuclear star clusters~\cite{2016ApJ...831..187A}, would then be more promising hosts.

\prlsec{Isolated or hierarchical origin?}
\label{ratesec}
We first investigate the likelihood of forming \event in either a hierarchical scenario or from isolated binaries in galactic fields.

To estimate the probability of forming \event in the latter case, we make use of publicly available population-synthesis distributions from Refs.~\cite{2018PhRvD..98h4036G,2019PhRvD..99j3004G} obtained with the {\sc StarTrack} \cite{2008ApJS..174..223B} and {\sc precession} \cite{2016PhRvD..93l4066G} codes. These are existing simulations realized with the setup of Ref.~\cite{2016A&A...594A..97B}, which have not been revisited or fine-tuned in any way to reproduce \event. In these simulations, supernova kicks have isotropic orientations and amplitudes drawn from a Maxwell distribution with one-dimensional dispersion $\sigma$ between 0 (implying that all BHs have strong fallback at formation) and 265 km/s (as estimated using proper-motion measurement of galactic pulsars \cite{2005MNRAS.360..974H}).
{\sc StarTrack} simulations provide masses and redshifts, as well as the evolution of the binary's orbital plane. This information is then used to add spins in postprocessing~\cite{2018PhRvD..98h4036G} and evolve them to the LIGO/Virgo band using {\sc precession}.
In particular, we use the ``uniform'' model of Ref.~\cite{2018PhRvD..98h4036G}, where the component spin magnitudes are distributed uniformly in $[0,1]$. This is a conservative assumption in this context, given the measured spins of  \event. We assume the spins to be initially aligned to the angular momentum of the binary, and we track the evolution of their orientations due to supernova kicks and tidal interactions (for tides, we use the ``time'' model of Ref.~\cite{2018PhRvD..98h4036G}).
Rates are computed using the standard noise curve of LIGO at design sensitivity \cite{2018LRR....21....3A}. We then filter the resulting synthetic catalogs looking for events that match
chirp mass $M_c$, mass ratio $q$, effective spin $\chi_{\rm eff}$, and precession parameter $\chi_{\rm p}$ of \event within the 90\% confidence intervals of their marginalized distributions. %
Because measurement correlations are neglected, this procedure selects a broader region of the parameter space compared to the actual support of the distributions, thus making our results conservative.

\begin{figure}[t]
  \includegraphics[width=\columnwidth,page=1]{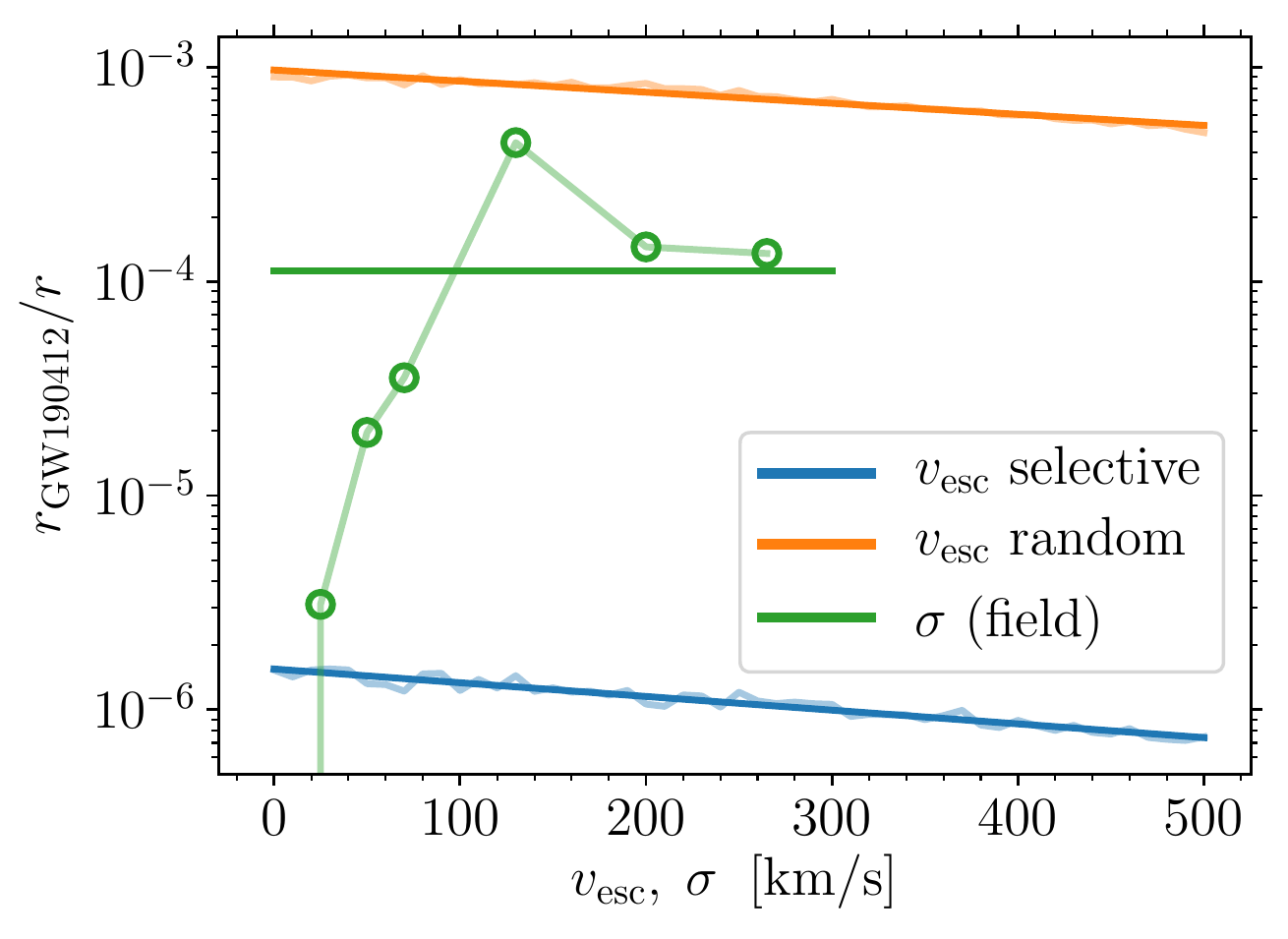}
  \caption{Fraction of the detection rate $r_{\rm \event}$ compatible with binaries similar to \event and the total detection rate $r$ predicted by each model. Blue and orange curves show results from the hierarchical formation model of  Ref.~\cite{2019PhRvD.100d1301G} as a function of the escape speed of the environment $v_{\rm esc}$. The blue curve (``selective'') assumes a population calibrated to the O1+O2 fit with the addition of second-generation mergers. The orange curve (``random'') assumes that all BHs in the environment pair with equal probability. Green circles are computed from existing population-synthesis simulations of isolated binary stars~\cite{2018PhRvD..98h4036G} (an additional data point at $\sigma=0$ and $r_{\rm \event}/r=0$ is not shown). In this case, rates are shown as a function of the strength $\sigma$ of the kicks imparted to BHs at birth. In all cases, thin lines connect results from the simulations, while the thicker line is a log-linear (constant) fit to the case of hierarchical (field) binaries.}
  \label{clusterrate}
\end{figure}

\begin{figure*}[t]
  \includegraphics[width=\textwidth,page=1]{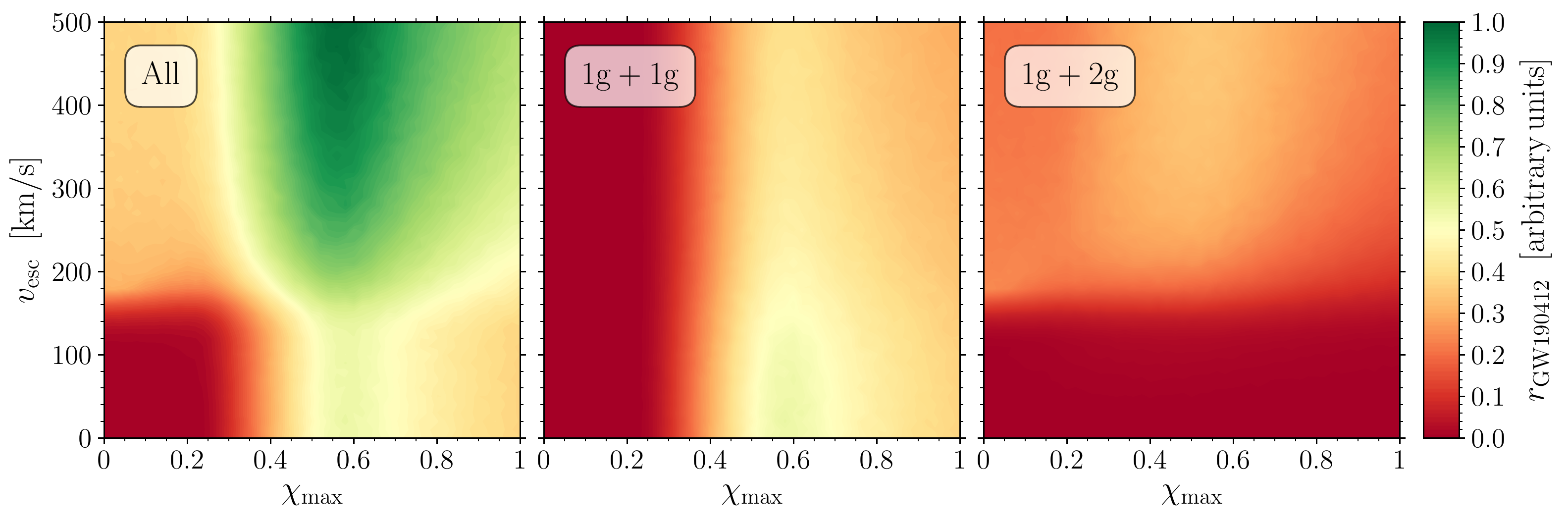}
  \caption{Constraints on the environment of \event in the hierarchical formation channel. We show the unnormalized contribution to the detection rate of binaries compatible with \event (color scale) as a function of the largest BH spins at birth $\chi_{\rm max}$ (x axis) and the escape speed $v_{\rm esc}$ of the environment, assuming that BHs pair randomly ($\alpha=\beta=0$). The left panel shows the entire population; the middle panel contains only mergers where both BHs are of first generation (1g+1g); the right panel contains the subset of events where a second-generation BH merges with a member of the injected first-generation population (1g+2g). Contributions from subpopulations of higher generations (2g+2g, 3g, etc) are subdominant.}\label{gridplot}
\end{figure*}

The green curve in Fig.~\ref{clusterrate} shows, as a function of $\sigma$, the fraction of the detection rate compatible with binaries similar to \event. This fraction is  $\mathcal{O}(10^{-4}$): even if all BH binaries were to originate from the isolated channel, GW190412 should appear only in a catalog with $\ssim 10^4$ entries. The fact that this event has been observed after $\ssim 50$ triggers is unusual for these models. The constraint in $q$ plays a dominant role, as it is responsible for the exclusion of the vast majority of the binaries from the original samples. Only $\ssim 1.5\%$ of the rate is accounted for by binaries with $0.2<q<0.4$, compared to $\ssim 80\%$ for sources with $q>0.8$. For $\sigma=0$, \event cannot be produced as
all BH spins are exactly aligned with the orbital angular momentum, which is incompatible with the measured value of $\chi_p$. Fluctuations as a function of $\sigma$ are likely due to low statistics of these preexisting simulations: \event is a rare event that belongs to the tail of the distributions.

We then perform a similar calculation using the simplified dynamical formation model of Ref.~\cite{2019PhRvD.100d1301G}. We consider an initial collection of $N=5000$ BHs with masses extracted from a power-law distribution $p(m) \propto m^{-2.3}$~\cite{2001MNRAS.322..231K}, spin magnitudes distributed uniformly in [0,1], and isotropic spin directions. These BHs are paired selectively to match the mass properties of the first ten BH events measured in O1 and O2~\cite{2019ApJ...882L..24A}, i.e. we set $p(m_1)\propto m_1^\alpha$, $p(m_2|m_1) \propto m_2^{\beta}$ with %
$\alpha=-1.6$ and $\beta = 6.7$. At each merger event, we estimate mass, spin, and recoil of the remnant using fitting formulas to numerical-relativity simulations (see Ref.~\cite{2019PhRvD.100d1301G} for details). Merger products are removed from the system if their recoil exceeds some escape speed $v_{\rm esc}$, which is a free parameter of the model. We then compute detection rates for LIGO at design sensitivity as in Ref.~\cite{2019PhRvD.100d1301G}, and we record the fraction of the total rate compatible with binaries similar to  GW190412.

In practice, this ``selective'' dynamical formation model assumes the BH population from O1 and O2, while also allowing for additional second-generation mergers.
The fraction of events compatible with \event is shown in Fig.~\ref{clusterrate} as a function of $v_{\rm esc}$ (blue curve). In this case, the fraction of the observable events that could form  GW190412 is $\ssim 10^{-6}$. The escape speed only changes these relative rates by a factor of a few, with lower (larger) $v_{\rm esc}$ corresponding to fewer (higher) second-generation events and higher (lower) rates for events like \event. Even allowing for second-generation mergers and moderate natal spins, the new event is an extremely unusual draw from the O1+O2 population.

To bracket the uncertainties, we then repeat the same exercise, but this time we pair the BHs in our sample randomly, i.e. we set $\alpha=\beta=0$ (solid orange curve in  Fig.~\ref{clusterrate}). In this case, the fraction of detectable sources compatible with \event is $\ssim 10^{-3}$. This is 3 orders of magnitude larger than the equivalent dynamical models with ``selective'' pairing, and 1 order of magnitude larger than the fraction of events compatible with \event for field binaries.

An important ingredient missing in Fig.~\ref{clusterrate} is the mixing fraction between the different formation channels. In other words, we compute the fraction of the rate compatible with \event \emph{within each model}, thus implicitly assuming that those scenarios are all equally probable. Without further assumptions, our analysis does not predict whether the isolated or dynamical formation channels are more likely to have formed \event. %
Our main message here is that both models struggle to reproduce the event: under all of our assumptions, \event appears to be extremely unusual, considering that the public O3 trigger list contains only $\ssim 50$ entries~\cite{GraceDB-O3-Public}. %

\prlsec{Astrophysical constraints.}
Let us now assume that \event was indeed formed as a second-generation merger.
What could we infer about its astrophysical environment?

Again, we make use of the ``random''  hierarchical formation model of Ref.~\cite{2019PhRvD.100d1301G} ($\alpha=\beta=0$), which presented the highest compatible rate compared to field binaries and other pairing prescriptions. Figure~\ref{gridplot} illustrates the likelihood of a given environment to be the birthplace of \event. We vary the escape speed $v_{\rm esc}$ and the largest natal spin $\chi_{\rm max}$ (i.e., first-generation BH spin magnitudes are extracted from a uniform distribution between 0 and $\chi_{\rm max}$). The parameter $\chi_{\rm max}$ encodes information about physical processes such as core-envelope angular-momentum transport and tidal interactions. As before, we select the mergers with observed properties within the 90\% confidence intervals of \event. The parameters of binaries that survive these cuts are, by construction, all very similar to each other, so their detection rate will also be approximately the same. Therefore the detector sensitivity and antenna patterns only enter the  overall rate normalization, which is not captured in Fig.~\ref{gridplot}.

The left panel of Fig.~\ref{gridplot} shows the entire population compatible with \event. The region with $\chi_{\rm max}\lesssim 0.3$ and $v_{\rm esc}\lesssim150$\,km/s is disfavored, while values of $\chi_{\rm max}\sim 0.6$ and $v_{\rm esc} \gtrsim 300$\,km/s  are preferred.  This follows from two complementary constraints:

\begin{itemize}
\item If one insists on explaining \event as a first-generation BH, then natal spins must allow for the measured value of $\chi_{\rm eff}\ssim 0.25$. The middle panel of Fig.~\ref{gridplot} shows the 1g+1g subset of the population of compatible binaries and it illustrates that, as expected, it is highly improbable to form \event if $\chi_{\rm max}\lesssim 0.3$.

\item BH remnants can merge again only if their gravitational recoil speed is smaller than $v_{\rm esc}$.  The right panel of Fig.~\ref{gridplot} shows the subset of events where one of the two binary members originated from a previous merger (1g+2g). One of the components of \event could be of second generation only in an environment with $v_{\rm esc}\gtrsim 150$ km/s. %

\end{itemize}

The threshold value $v_{\rm esc}\gtrsim 150$ km/s is set by the physics of gravitational recoils. In the $\chi_{\rm max}\to 0$ limit, all first-generation BHs are nonspinning and their recoils are bounded by $v_k\lesssim 175$ km/s~\cite{2007PhRvL..98i1101G}. Therefore, ejections of second-generation BHs can only take place in environments with escape speeds below this critical value. The same trend remains valid in the more general scenario where first-generation spins are nonzero: although much larger kicks are possible in this case, they are rare, and the vast majority of the BHs are imparted recoils of $\mathcal{O}(100\, {\rm km/s})$~\cite{2010ApJ...715.1006K,2012PhRvD..85l4049B,2012PhRvD..85h4015L,2018PhRvD..97j4049G}.

Together, these two constraints exclude the region of the parameter space that is perhaps better motivated astrophysically. If natal spins are as low as $\ssim 0.01$ \cite{2019ApJ...881L...1F}, it is highly unlikely to form \event in low-escape-speed environments like globular clusters~\cite{2013LRR....16....4B}.

\setlength{\tabcolsep}{3.5pt}
\renewcommand{\arraystretch}{1.5}
\begin{table}
\centering
\begin{tabular}{c|cc@{\hskip 10pt}||@{\hskip 10pt}c|cc}
& 1g+1g & 1g+2g && 1g+1g & 1g+2g\\
\hline
$M_{\rm c}$&
$13.29_{-0.54}^{+0.59}$
&
$13.26_{-0.46}^{+0.65}$
&
\chieff&
$0.21_{-0.15}^{+0.09}$
&
$0.22_{-0.15}^{+0.10}$
\\
$q$&
$0.32_{-0.08}^{+0.22}$
&
$0.29_{-0.08}^{+0.23}$
&
\chip&
$0.34_{-0.20}^{+0.33}$
&
$0.46_{-0.25}^{+0.23}$
\\
$m_1$&
$27.82_{-6.87}^{+5.32}$
&
$29.18_{-7.73}^{+5.84}$
&
$\chi_1$&
$0.42_{-0.30}^{+0.29}$
&
$0.56_{-0.21}^{+0.19}$
\\
$m_2$&
$8.94_{-1.23}^{+2.48}$
&
$8.56_{-1.10}^{+2.74}$
&
$\chi_2$&
$0.56_{-0.48}^{+0.39}$
&
$0.56_{-0.50}^{+0.39}$
\\
\end{tabular}
\caption{Medians and 90\% symmetric credible intervals for \event using both the standard prior (1g+1g) and an alternative prior where the primary component is assumed to come from a previous BH merger (1g+2g). Masses are reported in the source frame in solar masses.}
\label{90table}
\end{table}

\begin{figure}[t]
\includegraphics[width=\columnwidth]{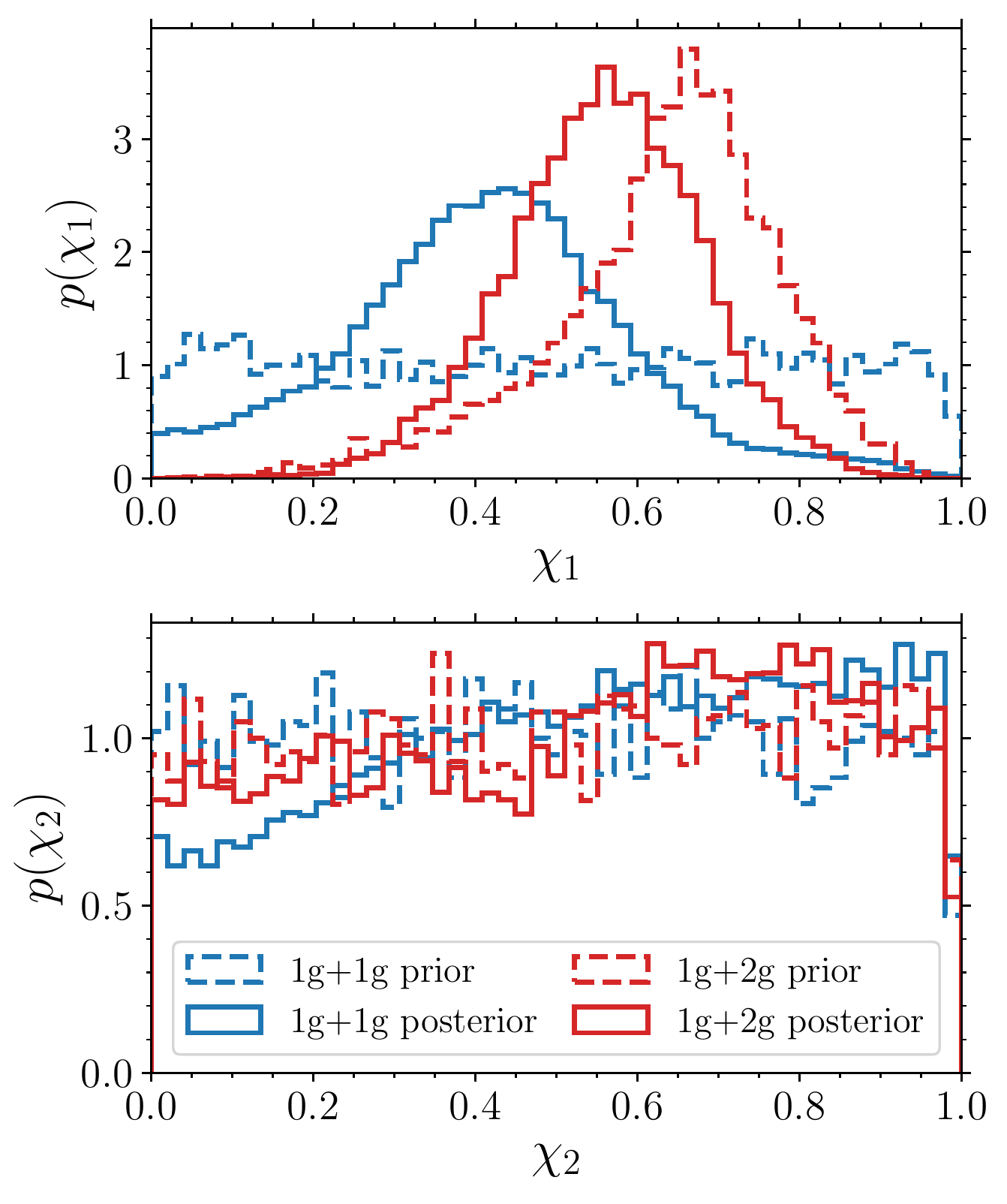}
  \caption{Marginalized prior (dashed) and posterior (solid) distributions of the spin magnitudes $\chi_1$ of the primary BH (top panel) and $\chi_2$ of the secondary BH (bottom panel), as obtained with first-generation (blue) and second-generation (red) priors.}\label{Fig.Post}
\end{figure}

\prlsec{A second-generation prior}
We now wish to verify whether LIGO/Virgo data for \event support a model where one of the two binary components is the result of a previous merger (1g+2g). We proceed by enforcing Bayesian priors tuned to plausible mass and spin distributions of second-generation mergers (see Refs.~\cite{2019PhRvD.100j4015C,2020RNAAS...4....2K} for complementary approaches applied to GW170729).  We use the \textsc{LALInference} source-characterization algorithm \cite{2015PhRvD..91d2003V} and public gravitational-wave strain data~\cite{2015JPhCS.610a2021V,gwosc}. %

We start with a ``first-generation prior'' consistent with the one used in Ref.~\cite{2020arXiv200408342T}: detector-frame component masses are distributed uniformly in $[3,\,50]~M_\odot$, spin magnitudes are distributed uniformly in $[0,0.99]$,  spin directions, orbital orientation and sky position are assumed to be isotropic, the luminosity distance is distributed uniformly in comoving volume in $[1,1200]$~Mpc, the phase at coalescence is distributed uniformly in $[0,2\pi]$, and the arrival time is distributed uniformly in $t_g\pm0.1$~s, where $t_g$ is the trigger time recorded by the search algorithm~\cite{GraceDB-O3-Public}.

We repeatedly draw pairs of random samples from this first-generation prior. Each pair is used to create a remnant BH with mass and spin estimated from the numerical-relativity fitting formulas of Refs.~\cite{2012ApJ...758...63B,2016ApJ...825L..19H}, as implemented in Ref.~\cite{2016PhRvD..93l4066G}.  The resulting joint distribution %
constitutes our two-dimensional prior for the mass and spin magnitude of second-generation BHs. We did not try to approximate our second-generation prior analytically, but rather augmented \textsc{LALInference} to use a numerical interpolant obtained from $5\times10^5$ previously stored evaluations.

Next, we analyze the data assuming that mass and spin magnitude of the more massive (primary) BH are extracted from the second-generation prior, whereas all other parameters (including the mass and spin magnitude of the secondary) are sampled from the first-generation prior described above.  We use the \textsc{IMRPhenomPv2} waveform model~\cite{2014PhRvL.113o1101H} %
and the publicly released estimates for the power spectral density~\cite{gwosc}. We do not marginalize over instrument calibration errors, but this does not significantly affect the inference on intrinsic parameters like masses and spins~\cite{2012PhRvD..85f4034V,2017PhRvD..96j2001C}.  We also perform a control run using the first-generation prior for both compact objects,
and verify that it %
mimics the results of Ref.~\cite{2020arXiv200408342T}.

Medians and $90\%$ symmetric credible intervals of the intrinsic parameters obtained with both sets of assumptions are listed in Table~\ref{90table}. Figure~\ref{Fig.Post} shows the one-dimensional marginalized priors and posteriors for the spin magnitudes.

The main parameters affected by the second-generation prior are the primary spin $\chi_1$ and the precession combination \chip (for $q \ll 1$ one has $\chi_1\simeq \sqrt{\smash\chieff^2+\smash\chip^2}$). Whereas the first-generation prior for $\chi_1$ was uniform, the second-generation prior peaks at $\chi_1\ssim 0.7$. This moves the posterior toward higher values relative to the first-generation analysis. Conversely, the spin of the secondary (which had a mild preference for large values in the first-generation analysis) is mostly flat using a second-generation prior. Altogether, this results in \chip peaking at larger values in the second-generation analysis, although part of that offset is driven by the prior. The second-generation \chieff  ($q$) posteriors peaks at slightly higher (lower) values, reflecting the anticorrelation between these two parameters~\cite{2018PhRvD..98h3007N}. The chirp mass posterior is unaffected.

When analyzing data with new priors, especially if narrower, it is important to calculate the Bayes factors $\mathcal{B}$ to check whether the new priors are disfavored by the data~\cite{2017PhRvL.119y1103V,2018IAUS..338...22G,2020ApJ...899L..17Z}. We report $\ln \mathcal{B}^{\rm 1g+1g}_{\rm 1g+2g}= 1.0$. While this implies that no strong conclusions can be drawn, it also suggests that the data do not significantly \emph{penalize} our narrower second-generation prior. This is far from obvious, as the second-generation prior excludes regions in the parameter space where the first-generation posterior had support.

Motivated by tidal spin-up in isolated binaries, Ref.~\cite{2020ApJ...895L..28M} reweighted the public \event posterior samples in favor of a prior where $\chi_1=0$ and only the secondary BH is spinning. This is very different from our assumptions, where the primary is a rapidly rotating remnant.

To keep the computational cost reasonable, we used a waveform model that does not include higher-order modes, which however are detectable in \event~\cite{2020arXiv200408342T}. We expect our key results, and in particular the absence of significant evidence in favor or against the second-generation model, to be unaffected by this choice, because higher harmonics mostly affect the estimation of extrinsic parameters of \event, such as the luminosity distance and orbital orientation of the binary. %

\prlsec{Conclusions.}
The detection of \event challenged previous  predictions. BH binaries with large spins and unequal masses cannot be easily accommodated in any of the major formation channels. Interpreting one component of \event as a remnant of a previous BH merger can reconcile the observation with small spins at birth while explaining the low value of its mass ratio. However, it also requires a dynamical environment with escape speed $\gtrsim 150$ km/s, thus excluding prominent hosts like globular and young star clusters. Analyzing \event using a Bayesian prior motivated by second-generation mergers returns Bayes factors  of order unity, implying that this assumption is 
equally supported by the data compared to scenarios where both BHs originate from stellar collapse.

\prlsec{Acknowledgments.}
We thank Vishal Baibhav, Thomas Callister, Roberto Cotesta, Nicola Giacobbo, Matthew Mould,
Carl Rodriguez, and Mike Zevin for discussions.
D.G. is supported by Leverhulme Trust Grant No. RPG-2019-350.
S.V. acknowledges the support of the  NSF and the LIGO Laboratory. LIGO was constructed by the California Institute of Technology and Massachusetts Institute of Technology  and operates under cooperative agreement NSF PHY-1764464.
E.B. is supported by NSF Grant Nos. PHY-1912550 and AST-2006538, NASA ATP Grants Nos. 17-ATP17-0225 and 19-ATP19-0051, and NSF-XSEDE Grant No. PHY-090003.
E.B. also acknowledges support from the Amaldi Research Center funded by the MIUR program ``Dipartimento di Eccellenza''~(CUP: B81I18001170001).
The authors would like to acknowledge networking support by the GWverse COST Action CA16104, ``Black holes, gravitational waves and fundamental physics.'' Computational work was performed on the University of Birmingham BlueBEAR cluster, the Athena cluster at HPC Midlands+ funded by EPSRC Grant No. EP/P020232/1, and the Maryland Advanced Research Computing Center (MARCC).
We are grateful for computational resources provided by Cardiff University  and funded by an STFC grant supporting UK Involvement in the Operation of Advanced
LIGO. 
The population-synthesis simulations used in this Letter are available at \href{https://github.com/dgerosa/spops}{github.com/dgerosa/spops}.
This research has made use of data, software and web tools obtained from the Gravitational Wave Open Science Center (\href{https://www.gw-openscience.org}{www.gw-openscience.org}), a service of LIGO Laboratory, the LIGO Scientific Collaboration, and the Virgo Collaboration. LIGO is funded by the U.S. National Science Foundation. Virgo is funded by the French Centre National de Recherche Scientifique (CNRS), the Italian Istituto Nazionale della Fisica Nucleare (INFN), and the Dutch Nikhef, with contributions by Polish and Hungarian institutes.
This is LIGO Document DCC-P2000161.

\bibliography{gw190412bib}

\end{document}